\documentclass[fleqn,twoside,11pt]{article}
\pdfoutput=1
\usepackage{amsfonts,amsmath,amssymb}
\usepackage{latexsym}
\usepackage[utf8]{inputenc}
\usepackage[english,russian]{babel}
\usepackage{verbatim}
\usepackage{cite}

\begin{document}

\twocolumn[
\title{{\bf On Geometrization of Classical Fields II\\  (MES: dark matter, energy)}}
\author{{\bf V. I. Noskov}}	       	
\date{\small \em Institute of Continuum Media Mechanics, Ural Branch of the Russian Academy of Sciences, Perm, Russia,
Email: nskv@icmm.ru}

\maketitle
{\bf Abstract --}{\small
The study of \cite{Nskv2023} geometrization of classical fields in the $4d$ Finsler space of MES (Model of Embedded Spaces) is continued. The model postulates a proper metric set of a distributed matter
{\em element} and states that the space-time of the Universe is a physical {\em Embedding} of such sets. The {\em Embedding} geometry is a Finsler-like {\em relativistic} geometry with connectivity
depending on the mechanical state of matter: torsion and non-metricity are absent. The Least Action Principle provides the geodesic motion of matter, leads to nonlinearity of the system of field equations,
anisotropy and Weyl-invariance of gravitation MES. It is shown that in the special case  of {\em Embedding} (conformal Weyl metric) the geometrization of fields can be realized {\em completely}: namely, to
obtain Maxwell-type equations and to find the gravitational sources, lying behind the $\lambda_c$ term of Einstein-type equation. They are identified as {\em dark} matter and energy of the Universe, and the
estimates of their material-field composition are close to the observed ones. It is also shown that the {\em Embedding's} gravitational and electromagnetic potentials are mutual gauge fields. Further development
of the MES implies going beyond the Weyl metric.}


{\bf Keywords:} {\it geometrization,\ gravity,\ electricity,\ Model of Embedded Spaces}
\bigskip
\medskip
]

\section{\textmd{INTRODUCTION}}
\label{intro}
\quad The relevance of geometrization of the classical electromagnetic field has always been undeniable, and recently, because of the latest discoveries in observational astronomy, it has gained special impor-\ tance.
By these findings we mean the accelerated expansion of the Universe \cite{Riess1998}, \cite{Perlmut1999} and peculiarities of the behavior of gravitating matter at the periphery of large cosmic objects like stellar
and galactic clusters.

Attempts to explain these phenomena from the standpoint of the classical theory of gravitation led to the hypothesis of {\em dark matter} that weakly participates in the electromagnetic interaction \cite{Blumenthal1984},
and {\em dark energy} \cite{Huterer2018} of the Universe, whose existence is confirmed by other indirect astronomi-\ cal measurements. At present these objects are compared to the cosmological term of Einstein's
gravitational equation  and at the phenomenologi-\ cal level are used in the theory of evolution of the Universe. Note that current models of the Universe are electrically neutral at all scales, from molar to cosmological,
and are governed only by the gravitational forces, while electromagnetism of the Universe is considered to be a local and weak property of its neutral structures.

On the other hand, the extreme saturation of the Universe with electromagnetic phenomena, the long-range character of the corresponding fields and the lack of reliable data on the electrical properties of the
Universe\footnote{A non-zero relic charge of the Universe is not excluded.} at all significant scales, do not exclude the electromagnetic generalization of the standard gravitational model. The generaliz-\ ed
model should contain distributed electrically charged matter and be controlled by classical fields: gravitational and electromagnetic. Obvious-\ ly, both gravitation and electromagnetism of the new model must
be geometrized, and the former should coincide in the electroneutral limit with gravitation of GR.

Finally, a history of attempts to geometrize electromagnetism \cite{Nskv2023} is also indicative of the need to replace the Einsteinian continuum with a diffe-\ rent (more adequate to the nature of electricity)
model of space-time.

The required replacement is the Model of Embedded Spaces (MES, {\em Embedding}), which has been proposed and used as a basis for a number of attempts of joint geometrization of gravitation and electricity
made in our early works. The intermediate result of these studies is quite comple-\ tely described in \cite{Nskv2023}. It turned out that the solution to the problem depends on the answer to some "eternal"\
physical questions and therefore is of essentially multivariant character.

In this paper we study the consistent and geometrically most reasonable variant of the MES-geometrization of (non-quantum) gravitation and electricity. Cosmological properties of the confor-\ mal Weyl's
{\em Embedding} (dark matter, dark energy) and internal symmetries of geometrized fields are discussed. The novelty of the MES also requires a brief description of its geometry.

\section{\textmd{ELEMENTS OF GEOMETRY}}
\label{elgm}
\quad The MES assumes that space-time is pseudo-Riemannian, 4-dimensional and a metric manifold called {\em Embedding} $M_e$. {\em Embedding} is an unstructu-\ red (like a mixture of liquids) {\em physical}
embedding of {\em proper} smooth point sets $M_c$ of the matter elements (each element of distributed matter has a spatial set  similar to $M_e$). The element interacti-\ ons keep {\em Embedding} from decaying and
determine the partial density of $M_c$ points of each element\footnote{Otherwise: a coordinate point of {\em Embedding} "contains"\ points of the proper $M_c$ the amount of which depends on the {\em interactions}
of its element.}

As in the original work \cite{Nskv2023}, {\em Embedding} is consi-\ dered in the {\em trial-particle approximation} and is the embedding of only two sets: the set of the Universe $M_s$ and the set $M_c$ of the {\em trial}
matter element\footnote{Set designation: $M_e, M_s, M_c$ is the manifold of Embedding, Source and Charge, respectively.}. In this approximation, the MES space is similar to a {\em fiber space} in he gauge gravitational
theories (Utiyama, Kibble, Konopleva, \cite{Konopl1972}) with $M_s$ as the {\em Base} and  $M_c$ as the {\em Layer}.

The metric tensor of {\em Embedding} depends on the mechanical state of an element of matter
\begin{gather}
 \label{1geo}
  \qquad\qquad\qquad g_{ik}=g_{ik}(x,u),\\
\label{2geo}
 \qquad u^i=dx^i/ds,\quad ds=+\sqrt{g_{ik}dx^idx^k},
\end{gather}
where $u^i$ is 4-velocity\footnote{In the general case $u^i$ is the unit tangent vector of the curve passing through the point $(x)$ \cite{Rund1959}.}. As an example, we can cite a conformal-Riemannian
metric of G. Weyl \eqref{1clw}.

The geodesic MES equation is the equation of matter motion  \cite{Nskv2023}
\begin{gather}
 \label{3geo}
  \frac{du_i}{ds}-\frac12u^ku^l\left(\frac{\partial}{\partial x^i}+2u^m\frac{\partial^2}{\partial x^{[i}\partial u^{m]}}\right)g_{kl}=0,
\end{gather}
where $\ldots_{[i,}\ldots_{k]}$ denotes antisymmetrization.

The geometry of {\em Embedding} is Riemannian and is constructed by the formal {\em substitution} of the normal gradient operator $\partial/\partial x^i$ for the generalized one
\begin{gather}
  \nonumber
  \qquad \partial/\partial x^i\quad \Rightarrow\quad \bar\partial/\partial x^i\equiv \partial/\partial x^i+\hat b_i,\\
  \nonumber
  \qquad\hat b_i=2u^k\hat b_{ik},\quad \hat b_{ik}=\partial^2/\partial x^{[i}\partial u^{k]},\\
  \label{4geo}
  \qquad \qquad \hat b_{[i} \hat b_{k]}=d\hat b_{ik}/ds,
 \end{gather}
where the gradients are "working"\ in {\em proper} sets:
\begin{gather}
    \label{4geoo}
   \qquad \partial/\partial x^i\in M_s,\quad \hat b_{ik}\in M_c,\quad \bar\partial/\partial x^i\in M_e,
\end{gather}
{\bf outside} which they are identically {\bf equal to zero}. Obviously, the operators $\hat b_i$ and $\hat b_{ik}$ are {\bf linear} ({\em first order} operators), since they define  increments {\em linear} in $dx^i$.

In work \cite{Nskv2023} we have chosen the simplest variant of the {\em Embedding} geometry: torsionless and without non-metricity. Its connectivity is the sum of the connectivity of the metric sets $M_s$ and $M_c$,
which can be demonstrated by generally covariant coordi-\ nate derivatives
\begin{gather}
  \nonumber
  \qquad A^i_{\ |k}\equiv \bar \partial A^i/\partial x^k + \Gamma^i\,_{lk}A^l\quad \Rightarrow\\
  \nonumber
  \qquad \qquad A^i_{\ |k}= A^i_{\ ;k}+A^i_{\ :k}, \\
  \nonumber
  \qquad A^i_{\ ;k}\equiv \partial A^i/\partial x^k+\gamma^i\,_{lk}A^l,\\
  \label{5geo}
  \qquad A^i_{\ :k}\equiv\hat b_k A^i+\omega^i\,_{lk}A^l,
\end{gather}
 where $A^i_{\ |k}$, $A^i_{\ ;k}$ and $A^i_{\ :k}$ are the derivatives in {\em Embedding} $M_e$, $M_s$ and $M_c$, respectively
 \begin{gather}
  \nonumber
  \qquad\qquad\Gamma^i\,_{kl}=\gamma^i\,_{kl}+\omega^i\,_{kl}, \\
  \label{6geo}
  2\Gamma_{i,kl}=\bar\partial g_{ik}/\partial x^l+\bar\partial g_{il}/\partial x^k-\bar\partial g_{kl}/\partial x^i, \\
  \nonumber
  2\gamma_{i,kl}=\partial g_{ik}/\partial x^l+\partial g_{il}/\partial x^k-\partial g_{kl}/\partial x^i,\\
  \label{7geo}
  2\omega_{i,kl}=\hat b_l g_{ik}+\hat b_k g_{il}-\hat b_i g_{kl}.
\end{gather}

Explicit expressions for curvatures of {\em Embedd-\ ing}\footnote{Analogous to the Riemannian case, the curvatures are general covariant scalars of the transformations \eqref{1ge}, \eqref{8ge}.} are
\begin{gather}
  \nonumber
  \qquad\qquad\bar R=R+r+p,\quad R=g^{ik}R_{ik}, \\
  \label{8geo}
  R_{ik}=\frac{\partial \gamma^l_{\ ik}}{\partial x^l}-\frac{\partial \gamma^l_{\ il}}{\partial x^k}+\gamma^l_{\ ik}\gamma^m_{\ lm}-\gamma^l_{\ im}\gamma^m_{\ kl}, \\
  \nonumber
  R=\frac{\partial \gamma^{l,m}_{\ \ \ m}}{\partial x^l}-\frac{\partial \gamma^{l,m}_{\ \ \ l}}{\partial x^m}+\gamma^{l,mn}\gamma_{n,ml}-\gamma^l_{\ ml}\gamma^{n,m}_{\ \ \ \ n},\\
  \nonumber
  r=4u^i\;\partial f_{i\ \ \ \ l}^{\;[k,l]}/\partial x^k-\xi_{ik}u^iu^k,\\
  \nonumber
  \xi_{ik}=f_{il,m}^{\ \ \ m}f_{k\ \ \ n}^{\ l,n}+f_{il,mn}\left(f_k^{\ l,mn}-2f_k^{\ m,ln}\right),\\
  \label{9geo}
  f_{ik,lm}\equiv\hat b_{ik}g_{lm}.
\end{gather}
The intersecting curvature is
\begin{gather}
  \nonumber
  p=2\left(\partial\omega^{[i,k]}_{\ \ \ k]}/\partial x^{[i}+\hat b_{[i}\gamma^{[i,k]}_{\ \ \ k]}+\right. \\
  \label{10geo}
  \left. \qquad\qquad\qquad+\omega_{i,kl}\gamma^{(k,l)i}-\omega^i_{\ ik}\gamma_l^{\ lk}\right)\equiv 0,
\end{gather}
since {\bf outside} its set the $\partial/\partial x^i\equiv 0$ and $\hat b_i\equiv 0$, as well as $\omega\cdot\gamma=0$.

\section{\textmd{COORDINATE AND PHYSICAL STATE TRANSFORMATIONS}}
\label{cootrans}
\quad The notions of "scalar"\!, "vector"\!, "tensor"\ and "connectivity"\ in the case of a singular (Riemanni-\ an) manifold are well known. The case of {\em Embedd-\ ing} obviously needs clarification. An arbitrary
local transformation of {\em Embedding's}  coordinate system $\{x^i\}$ to the system $\{x'^k\}$ is accompanied by the corresponding transformation of the veloci-\ ty of matter $u\rightarrow u'$ that is equivalent
to the transition of the physical system from the state $(x^i,u^k)$ to the close state $(x'^l,u'^m)$. By virtue of the {\em Embedding} structure (at least two, relatively moving sets $M_s$ and $M_c$) the transition $(x^i,u^k)\rightarrow (x'^l,u'^m)$ is described by anisotropic (in contrast to the isotropic Riemannian case) smooth continu-\ ous functions
\begin{gather}
 \label{1ge}
 \quad x^i=x^i(x'^k,u'^l),\quad u^i=u^i(x'^k,u'^l).
\end{gather}
In this case, since the coordinate transformation $x^i=x^i(x'^k,u'^l)$ defines the concept of "tensor"\!, it is the main one, determining the law of transforma-\ tion of the velocity vector $u^i=u^i(x'^k,u'^l)$.

{\em Embedding's} small increment operator (differ-\ ential)  is defined according to \eqref{4geo} as
\begin{gather}
 \label{2ge}
 \qquad\qquad\qquad \bar d=dx^i\,\frac{\bar\partial}{\partial x^i}.
\end{gather}
Then the differential of $\bar dx^i$ coordinate, being {\em infinitesimal vector}, is transformed according to
\begin{gather}
 \nonumber
 \bar dx^i=dx'^k\,\frac{\bar\partial x^i}{\partial x'^k}=\left(\frac{\partial x^i}{\partial x'^k}+2u'^l\frac{\partial^2 x^i}
 {\partial x'^{[k}\partial u'^{l]}}\right)\, \bar dx'^k,
 \end{gather}
since $\bar dx'^k=dx'^k$ (as well as $\bar dx^i=dx^i$)\footnote{$\bar dx^i=dx^k(\partial/\partial x^k+\hat b_k)x^i\simeq dx^i$ -- in the case of {\em differential}, the orthogonal summand can be neglected
considering the latter as a correction of the next order of smallness. This cannot be done for an arbitrary vector.} are {\em generally covariance vectors}. Hence, an arbitrary contravariant vector $A^i$ of
{\em Embedding} is transfor-\ med in a similar way
\begin{gather}
 \label{3ge}
 A^i=\left(\frac{\partial x^i}{\partial x'^k}+\hat b'_{\,k} x^i\right)A'^k\equiv\left(\bar\partial x^i/\partial x'^k\right)A'^k.
\end{gather}
It can be seen that the transformation matrix consists of two summands, the first of which is isotropic Riemannian (contribution of the set $M_s$ -- longitudinal transfer), and the second is anisotropic (contribution of
the set $M_c$ -- the transfer orthogonal to the displacement vector $dx$).

The law of transformation of a covariant vector follows from the condition of invariance of the scalar product $A^iB_i=const$ and is formally written as
\begin{gather}
 \label{4ge}
\qquad\qquad A_i=\left(\partial x'^k/\bar\partial x^i\right)\,A'_k
\end{gather}
where the transformation matrix $\partial x'^k/\bar\partial x^i$, being the inverse of the matrix $\bar\partial x^i/\partial x'^k$ \eqref{3ge}, can be found by the rules of matrix algebra.

The differential of {\em Embedding's} scalar function $\bar d\varphi$ being a scalar, i.e. an invariant of transfor-\ mations \eqref{1ge}, can be represented according to \eqref{2ge} and the fact that
$dx^i=\bar dx^i$, as a scalar product of two {\em Embedding's} vectors
\begin{gather*}
 \qquad\qquad \bar d\varphi=dx^i\,\frac{\bar\partial\varphi}{\partial x^i}=\bar dx^i\,\frac{\bar\partial\varphi}{\partial x^i},
\end{gather*}
from which it follows that the common covariant vector $\bar\partial\varphi/\partial x^i$ is transformed according to \eqref{4ge} as
\begin{gather}
 \nonumber
 \qquad\qquad \frac{\bar\partial\varphi}{\partial x^i}=\left(\partial x'^k/\bar\partial x^i\right)\,\frac{\bar\partial\varphi}{\partial x'^k}\quad \Rightarrow \\
\label{5ge}
\qquad\qquad \frac{\bar\partial}{\partial x^i}=\left(\partial x'^k/\bar\partial x^i\right)\,\frac{\bar\partial}{\partial x'^k}.
\end{gather}

The noncontradiction of the "vector"\ definition \eqref{3ge}, can be demonstrated by the  law of transfor-\ mation of the velocity vector $u^i$. According to \eqref{3ge} we have
\begin{gather}
\label{6ge}
\qquad\qquad\qquad  u^i=\left(\partial x^i/\partial x'^k\right)u'^k
\end{gather}
since $u'^k\,b'_k=0$ (see also footnote$^6$ page 3). The transformation matrix $\partial x^i/\partial x'^k$ is exactly the local Lorentzian boost, which is {\em admissible} for $u^i$. The obtained law "cuts down"\
the velocity transformation matrix \eqref{1ge} to a matrix depend-\ ing only on the coordinate.

The law of transformation of the metric as a state-dependent function $(x,u)$ is determined by matrices that also depend on the state (initial or final -- depending on the covariance). For example,
\begin{gather}
  \nonumber
  g_{ik}(x,u)=\left(\partial x'^l/\bar \partial x^i\right)\left(\partial x'^m/\bar \partial x^k\right)g'_{lm}(x',u'),\\
  \label{7ge}
  g^{ik}(x,u)=\left(\bar \partial x^i/\partial x'^l\right)\left(\bar \partial x^k/\partial x'^m\right)g'^{lm}(x',u').
\end{gather}

Finally, let us consider the law of transforma-\ tion of {\em Embedding's}, $M_s$ and $M_c$ connectivities, using as an example the transformations of an arbitrary tensor $A^i_{\ |k}$.  According to definitions
\eqref{5geo} and \eqref{3geo} we have
\begin{gather*}
    \qquad A^i_{\ |k}\equiv \bar \partial A^i/\partial x^k + \Gamma^i\,_{lk}A^l=\\
   =\frac{\bar\partial}{\partial x^k}\left(\frac{\bar\partial x^i}{\partial x'^l}A'^l\right)+\Gamma^i\,_{lk}\left(\frac{\bar\partial x^l}{\partial x'^m}A'^m\right)=\\
   =\frac{\bar\partial x^i}{\partial x'^l}\frac{\partial x'^m}{\bar\partial x^k}\left(\frac{\bar\partial A'^l}{\partial x'^m}\pm \Gamma'^l\,_{nm}A'^n\right)+\\
   +\left(\frac{\bar\partial x^m}{\partial x'^l}\Gamma^i\,_{mk}+\frac{\partial x'^m}{\bar\partial x^k}\frac{\bar\partial^2 x^i}{\partial x'^m\partial x'^l}\right)A'^l.
\end{gather*}
Hence it immediately follows that
\begin{gather*}
  \frac{\bar\partial x^m}{\partial x'^n}\Gamma^i\,_{mk}=\frac{\bar\partial x^i}{\partial x'^p}\frac{\partial x'^m}{\bar\partial x^k}\Gamma'^p\,_{nm}-
  \frac{\partial x'^m}{\bar\partial x^k}\frac{\bar\partial^2 x^i}{\partial x'^m\partial x'^n}
\end{gather*}
and finally we have
\begin{gather}
  \nonumber
  \qquad\qquad\Gamma^i\,_{lk}=\frac{\bar\partial x^i}{\partial x'^p}\left[\frac{\partial x'^n}{\bar\partial x^l}\frac{\partial x'^m}{\bar\partial x^k}\Gamma'^p\,_{nm}+\right.\\
  \nonumber
  \qquad\qquad\qquad\left.+\frac{\partial x'^m}{\bar\partial x^k}\frac{\bar\partial}{\partial x'^m}\left(\frac{\partial x'^p}{\bar\partial x^l}\right)\right] \Rightarrow\\
  \label{8ge}
  \Gamma^i\,_{kl}=\frac{\bar\partial x^i}{\partial x'^p}\left(\frac{\partial x'^m}{\bar\partial x^k}\frac{\partial x'^n}{\bar\partial x^l}\Gamma'^p\,_{mn}+
  \frac{\partial^2 x'^p}{\bar\partial x^k\bar\partial x^l}\right)
\end{gather}
to the next order of smallness in the second summ-\ and (just as in the limiting case $\bar dx^i=dx^i$, see footnote$^6$ page 3).

It is seen that the law \eqref{8ge} of {\em Embedding's} connectivity transformation is indeed Riemannian in the form. With account of the existence regions of the operators $\partial/\partial x^i\in M_s$
and $\hat b_i\in M_c$, it decomposes into analogous laws of transformation of connectivities $\gamma^i\,_{kl}$ and $\omega^i\,_{kl}$:
\begin{gather}
  \label{9ge}
  \gamma^i\,_{kl}=\frac{\partial x^i}{\partial x'^p}\left(\frac{\partial x'^m}{\partial x^k}\frac{\partial x'^n}{\partial x^l}\gamma'^p\,_{mn}+
  \frac{\partial^2 x'^p}{\partial x^k\partial x^l}\right),\\
  \nonumber
  \omega^i\,_{kl}=\hat b'\!_p x^i\left[\left(\hat b^{-1}\!_k x'^m\right)\left(\hat b^{-1}\!_l x'^n\right) \omega'^p\,_{mn}+\right.\\
  \label{10ge}
  \qquad\qquad\qquad\qquad\qquad \left.+\hat b^{-1}\!_k\left(\hat b^{-1}\!_l x'^p\right)\right].
\end{gather}

Note that the quantity $f_{ik,lm}$, like the connecti-\ vity $\omega_{i,kl}$, {\em is not a tensor} for the same reason -- the existence of the second derivative of the transformati-\ on matrix:
\begin{gather}
  \nonumber
  f_{ik,lm}=\left(\hat b^{-1}\!_i x'^n\right)\left(\hat b^{-1}\!_k x'^p\right)\left(\hat b^{-1}\!_l x'^r\right)\left(\hat b^{-1}\!_m x'^s\right)\cdot\\
 \label{11ge}
 \qquad\cdot\, f'_{np,rs}+2\hat b^{-1}\!_l x'^r\hat b^{-1}\!_m x'^s g'_{rs}\hat b_{ik}\hat b^{-1}\!_m x'^s.
\end{gather}

Furthermore, the coordinate derivatives by de-\ finition are taken {\em everywhere}, including transfor-\ mation matrices,  at $u=const$ (at a fixed direction in the locus). Therefore, in particular, the gravity of
{\em Embedding} is a generally-covariant phenomenon (as in GR).

In the general case transformations  \eqref{1ge} descri-\ be not only the Lorentz-Riemannian local boosts but also small translations. Therefore, they can be considered as local transformations of the states $(x,u)$
of the system.

Finally, the foregoing proves the possibility of constructing geometry and covariant tensor analysis both in the original sets and in {\em Embedding}.

\section{\textmd{LAGRANGE ACTION: UNDE-\\ TERMINED MULTIPLIERS AND EXTREMA}}
\label{actsys}
\quad The construction of {\em Embedding's} Lagrange acti-\ on for the physical system under consideration, immediately requires an answer to the two funda-\ mental questions: about the field hypothesis of mass
and about the point structure of matter.

The problem of the proper electromagnetic mass of a charge was intensively investigated in the 30s-40s of the last century \cite{Ivanenko1951}, \cite{LL1988}. The outcome is well known: the contribution of the proper electromagne-\ tic field to its mass is {\em zero}\footnote{The conclusion is valid both for the Maxwell theory and for the nonlinear ones of G. Mie (1912), M. Born (1934).}. This result was formally applied to both the point
particle model and the classical gravity. The more recent discovery of the Higgs boson \cite{Higgs1964} confirmed the correctness of the postulate: it is currently believed that the mass of matter is due to the interaction
with the Higgs field and does not have any relation to the proper electromagnetic and gravity fields.

Relativism limits the continuous distributions $\rho,\mu$ of charge and mass $(q,m)$ of classical matter
\begin{gather*}
\qquad\rho=dq/\sqrt{\gamma}\, dV,\quad \mu=dm/\sqrt{\gamma}\, dV,
\end{gather*}
where $\gamma=\|\gamma_{\alpha\beta}\|$ is the $3d$ determinant of the metric tensor $\gamma_{\alpha\beta}=-g_{\alpha\beta}+g_{0\alpha\beta}/g_{00}$, and $dV=dxdydz$ is the element of the $3d$ volume,
by {\em the particle point condition}
\begin{gather}
 \label{000as}
 \qquad\rho/q=\mu/m\quad \Rightarrow\quad  \beta\equiv \rho/\mu=q/m
\end{gather}
taken for a single particle. Therefore, for matter consisting of particles of the same kind $\beta=const (x)$, and in the general case the ratio of charge and mass densities of distributed matter is
\begin{gather}
  \label{00as}
  \qquad\qquad\beta=\beta(x),\quad \beta(x)\neq \beta(g_{ik}),
\end{gather}
(charge and mass are {\em equally distributed} \eqref{000as}).

Then the simplest action of  {\em Embedding's} physi-\ cal system must consist of at least two summands: the Lagrange density of {\em free} matter
\begin{gather}
  \label{0as}
  \qquad\qquad \Lambda_0=-\mu c^2\,ds/\sqrt{g_{00}}\,dx^0
\end{gather}
and the field Lagrange density $\sim\bar R/\varkappa$, $\bar R=R+r$, \eqref{8geo}-\eqref{9geo}. Obviously, the weight of the gravitational term ($\sim R$) must be proportional to $k^{-1}$, while the
weight of the electric term ($\sim r$) must be proportional to $\beta^{-2}$ and both values must belong to a {\em single} function. From thise follows the condition
\begin{gather}
\label{001as}
 \qquad\qquad\quad  \lim_{\rho\rightarrow 0} \beta^2=k,
\end{gather}
which combines both  the geometrical and physical constraints on the field Lagrangian density $\sim\bar R$.

Let us summarize the initial features of the developed geometrization. First, the field Lagrange densities $\sim R$ and $\sim r$ are taken with the {\em same} sign (unified geometrical nature), which is also
a {\em more} physically correct choice (as will become obvious in the future) than the choice in \cite{Nskv2023}, corresponding to the standard field one. Secondly, using the scalar $\bar R$ does not cause
objections for the same reasons as the use of $R$ in GR: $\bar R$ is the only geometrically scalar invariant of arbitrary transformations \eqref{1ge}, which containes the connectivity quadratic  summands. Thirdly,
condition \eqref{001as}, providing the {\em unified in the mean-\ ing} weighting factor for {\em different} field densities, applies well to both gravity and electricity. Finally, the {\em “electrovacuum”\ }
quantity $\beta_{0}=\pm \sqrt k$ of {\em Embedding's} matter can be experimentally verified \cite{Nskv2017}.

An obvious geometric choice of variational variables in \cite{Nskv2023}: “The independent variational para-\ meters to be chosen are the metric (at a fixed direction), $g^{ik}(x,u=const)$, and this direction itself at a
fixed point of {\em Embedding} -- the velocity of matter $u^i(x)$..."\ , adds the other {\em two}, equal to zero, but essentially important summands. (The latters are due to the normalization of the chosen variables
and the only {\em appropriate} variational method: {\em method of undetermined Lagrange multip-\ liers} \cite{Smirnov1956}.)

In view of the above statements we have
\begin{gather}
 \nonumber
 cS=\int\left[\Lambda_0-\left(R/\varkappa_g+r/\varkappa_e\right)/2+\right. \\
 \label{1as}
\left.+\lambda_g(g_{ik}g^{ik}-4)+\lambda_u(u_iu^i-1)\right]\sqrt{-g}d\Omega,
\end{gather}
where $\lambda_g$ and $\lambda_u$ are the {\em undetermined  multipliers} and the "field"\ weight constants are
\begin{gather}
 \label{2as}
  \qquad\qquad\varkappa_g=8\pi c^{-4}k,\quad \varkappa_e=8\pi c^{-4}\beta^2.
\end{gather}

The variational variables define two geometric extrema of the action: the extremum of {\em Embedd-\ ing's} two-dimensional point density at a fixed direction and the directional extremum at a fixed point density
\begin{gather*}
  \delta S[g^{ik},u^l]_{|u^l=const}=0,\quad   \delta S[g^{ik},u^l]_{|g^{ik}=const}=0,
\end{gather*}
giving the analogs of the Einstein and the Maxwell field equations\footnote{The method consists of three steps: (1) by varying \eqref{1as} we find the equations with {\em nonvariable and unknowns} $\lambda_g$ and
$\lambda_u$. The variation in $g^{ik}$ is taken under the assumption that $g_{lm}$ is {\em independent} ($\delta g_{mn}=0$). Similarly, we assume that $\delta u_p=0$  as it is {\em independent} of $u^l$.
That is, we {\em neglect} the normalizations (holonomy relations); (2) by using the {\em method-independent} considerations, we find $\lambda_g$ and $\lambda_u$; (3) {\em recovering} the normalizations in
the final results.} In the following, for the sake of brevity, we'll call them as the {\em gravity} and {\em electricity} equations of {\em Embedding}.

\section{\textmd{FIELD EQUATIONS}}
\label{eqfild}
\subsection{\textmd{Equation of electricity}}
\label{eqel}
\quad The sought equation is found by variation of the action\eqref{1as}, $\delta S[g^{ik},u^l]_{|g^{ik}=const}=0$, along the direction $u^i$ (from explicit dependence) at a fixed point density $M_e$
\begin{gather}
   \nonumber
   \qquad\qquad\frac{\delta \Lambda_0}{\delta u^i}-\frac{1}{2\varkappa_e}\frac{\delta r}{\delta u^i}+\lambda_u u_i=0\quad \Rightarrow\\
   \label{1ele}
   \qquad\qquad\delta r/\delta u^i=\varkappa_e\left(\Lambda_0+2\lambda_u\right)u_i
\end{gather}
since $\delta\Lambda_0/\delta u^i=\Lambda_0 u_i/2$.

Proof:
\begin{gather*}
 \frac{\delta\Lambda_0}{\delta u^i}=-\frac{\delta}{\delta u^i}\left(\frac{d(mc^2)}{\sqrt{-g}\,d\Omega}\,ds\right)=-\frac{d(mc^2)}{\sqrt{-g}\,d\Omega}\,\frac{\delta ds}{\delta u^i}=
 \end{gather*}
since the point density of {\em Embedding} is fixed, $\sqrt{-g}=const$ is the variation constant. Extracting the {\em explicit} dependence of $ds$ on $u^i$ yields
\begin{gather*}
 =\frac{\Lambda_0}{2ds^2}\frac{\delta(dx_kdx^k)}{\delta u^i}\equiv \frac{\Lambda_0}{2ds^2}\ ds^2\frac{\delta(u_ku^k)}{\delta u^i}=\frac12 \Lambda_0 u_i,
 \end{gather*}
because $u_k=const$ of variation, $\Box$.

Taking into account \eqref{9geo} we obtained the follo-\ wing  form of equation \eqref{1ele}\footnote{Equation for $g_{ik}(x=const,u)$ of {\em Embedding}.}
\begin{gather}
   \label{2ele}
   \quad 2\,\frac{\partial f_{i \ \ \ \ \  l}^{\ [k,l]}}{\partial x^k}=-\frac{\varkappa_e}{2}\left(-\Lambda_0g_i^{\ k}+a_i^{\ k}\right)u_k, \\
  \label{3ele}
   \qquad a_{ik}=(2/\varkappa_e)\left(-\xi_{ik}-\varkappa_e \lambda_u\, g_{ik}\right).
\end{gather}
It is seen that it contains an undetermined multiplier $\lambda_u$, the choice of which in the form
\begin{gather}
 \label{4ele}
  \qquad\qquad\qquad\lambda_u\sim -\xi/4\varkappa_e
\end{gather}
makes $a_{ik}$ similar to Maxwell EMT (Energy-Momen-\ tum Tensor) of electromagnetic field.

{\em However, there is no reason to suppose that the undetermined multiplier $\lambda_u$ depends only on the scalar $\xi$. It can also depend on the gravitational curvature $R$ of Embedding.} In order to clarify
this circumstance, we need to derive the gravity equation.

\subsection{\textmd{The equation of gravity}}
\label{eqgr}
\quad Homogeneity of $r$ with respect to $u^i$
\begin{gather*}
  u^i(\delta r/\delta u^i)=r-\xi_{ik}u^iu^k
\end{gather*}
and the \eqref{1ele} equation allow us to eliminate the senior derivative of $f_{ik,lm}$ from $r$:
\begin{gather}
  \label{1gre}
  r=\xi_{ik}u^iu^k+\varkappa_e(\Lambda_0+2\lambda_u)u^2.
\end{gather}

Substituting this relation into \eqref{1as} provides consideration for the extremum found above (by $u^i$)
\begin{gather}
 \nonumber
  2cS=\int\left[\Lambda_0-R/\varkappa_g-\xi_{ik}u^iu^k/\varkappa_e+\right.\\
 \label{2gre}
  \left.\qquad\qquad+2\lambda_g(g_{ik}g^{ik}-4)-2\lambda_u\right]\sqrt{-g}d\Omega,
\end{gather}
and the form of the obtained expression indicates that the correct relation between matter and gravita-\ tion in the subintegral expression can be achieved only at $-2\lambda_u\sim R/2\varkappa_g$. Then given
\eqref{4ele}, the {\em simplest choice} of the sought multiplier $\lambda_u$ such as
\begin{equation}
 \label{3gre}
  \qquad\qquad \lambda_u=-(\xi/\varkappa_e+R/\varkappa_g)/4,
\end{equation}
gives\footnote{On the gravitational sources of the MES: the condition $\lambda_u=const(u)$ of the variational procedure {\em admits} the appearance of an additional density in the action $S$
$\Lambda_{\not\, 0}(x,u)$, but {\em only} such that $\Lambda_{\not\, 0}\sim \Lambda_0$.} a new expression for functional \eqref{2gre}:
\begin{gather}
  \nonumber
 2cS=\int\!\!\left\{\!\Lambda_0-\!\left[\!R/\varkappa_g+\!\left(2\xi_{ik}u^iu^k-\xi\right)\!\!/\varkappa_e\right]\!/2\, +\right. \\
  \label{4gre}
  \left. \qquad\qquad\qquad +2\lambda_g\left(g_{ik}g^{ik}-4\right)\right\}\sqrt{-g}d\Omega.
\end{gather}

The action modified in such a way already allows us to search for a gravita-\ tional extremum: $\delta S[g^{ik},u^l]_{|u^l=const}=0$.
Varying \eqref{4gre} in $g^{ik}$ at $u^i=const$ we obtain the anisotropic gravity equation\footnote{Equation for $g_{ik}(x,u=const)$ of {\em Embedding}.}
\begin{gather}
\nonumber
  R_{ik}-g_{ik}R/2-4\varkappa_g \lambda_g g_{ik}=\\
 \label{5gre}
 \qquad\qquad=\varkappa_g\left(t^{(0)}_{\ \ \ ik}+t^{(oem)}_{\ \ \ \ ik}+t^{(fem)}_{\ \ \ \ ik}\right),
\end{gather}
where the EMT of free matter is
\begin{gather}
 \nonumber
  t^{(0)}_{\ \ \ ik}=2\,\frac{\partial\left(\sqrt{-g}\,\Lambda_0\right)}{\sqrt{-g}\,\partial g^{ik}}=-\frac{2c^2}{\sqrt{-g}}\frac{\partial}{\partial g^{ik}}\left(\frac{dm}{d\Omega} ds\right)=\\
 \nonumber
  =-\frac{2c^2}{\sqrt{-g}}\frac{dm}{d\Omega}\frac{\partial ds}{\partial g^{ik}}=\frac{\mu c^2}{\sqrt{g_{00}}\,dx^0}\frac{dx_idx_k}{ds}\quad \Rightarrow\\
 \label{6gre}
  \qquad t^{(0)}_{\ \ \ ik}=-\Lambda_0u_iu_k,\quad  t^{(0)}=-\Lambda_0>0;
\end{gather}
\quad the EMT of the {\em own electrometric} field of charged matter (convolution with velocities of motion) is\footnote{Here $\partial(u^lu^m)/\partial g^{ik}=0$ since $\delta u^i=0$.}
\begin{gather}
  \nonumber
t^{(oem)}_{\ \ \ \ ik}=-u^lu^m\left(2\partial/\partial g^{ik}-g_{ik}\right)\xi_{lm}/\varkappa_e\quad \Rightarrow \\
  \nonumber
\quad  \varkappa_e t^{(oem)}_{\ \ \ \ ik}=\left\{g_{ik}\xi_{lm}-4\left[f_{ln,ik}f_{m \ \ \ \ p}^{\ \ n,p}+f_{li,n}^{\ \ \ (n}\cdot \right.\right.\\
  \nonumber
\quad  \left.\left. \cdot f_{mk,p}^{\ \ \ \ \ p)}+2f_{ln,pi}\left(f_{m\ \ \ \ \ k}^{\ \ [n,p]}+f_{km,}^{\ \ \ \ np}\right)\right]\right\}u^lu^m, \\
  \label{7gre}
\qquad\qquad\qquad  \varkappa_e t^{(oem)}=-2\xi_{ik}u^iu^k;
\end{gather}
\quad and the EMT of the {\em free electrometric} field\footnote{The signs of $t^{(oem)}_{\\ \ \ ik}$ and $t^{(fem)}_{\ \ \ \ ik}$ are obviously opposite to the \cite{Nskv2023} due to the choice
of \eqref{1as}. }
\begin{gather}
  \nonumber
t^{(fem)}_{\ \ \ \ ik}=\left(2\partial/\partial g^{ik}-g_{ik}\right)\xi/2\varkappa_e\quad \Rightarrow \\
  \nonumber
\quad  \varkappa_e t^{(fem)}_{\ \ \ \ ik}=\xi_{ik}-\frac{g_{ik}}{2}\xi+2\left[f_{lm,ik}f^{lm,n}_{\ \ \ \ \ n}+ \right. \\
  \nonumber
\quad  \left.  +f_{il,m}^{\ \ \ (m}f_{k \ \ \ n}^{\ l,n)}+2f_{lm,ni}\left(f_{\ \ \ \ \ \ k}^{l[m,n]}+f_k^{\ l,mn}\right)\right], \\
  \label{8gre}
\qquad\qquad\qquad  \varkappa_e t^{(fem)}=2\xi.
 \end{gather}

Substituting \eqref{3gre} into \eqref{3ele} gives finally
\begin{gather}
 \nonumber
 \quad a_{ik}=\tau_{ik}+\frac{R}{2\varkappa_g}\,g_{ik},\quad \tau_{ik}\equiv \frac{2}{\varkappa_e}\left(-\xi_{ik}+\frac{g_{ik}}{4}\xi\right), \\
 \label{9gre}
 \qquad\qquad\qquad a=2R/\varkappa_g, \quad \tau=0, \\
 \label{10gre}
 2\frac{\partial f_{i \ \ \ \  l}^{\, [k,l]}}{\partial x^k}=\!-\frac{\varkappa_e}{2}\!\left[\!\left(\!-\Lambda_0+\!\frac{R}{2\varkappa_g}\right)\!g_i^{\ k}\!+\!\tau_i^{\ k}\right]\!u_k.
\end{gather}

It would seem that tensors \eqref{9gre} completely define the electricity equation and its derivation has been finished. However, that is not the case: as seen from the gravitational equation \eqref{5gre}, the curvature
$R$ depends on the multiplier $\lambda_g$,
\begin{gather}
 \label{11gre}
 R/2\varkappa_g=\!-(t^{(0)}+t^{(oem)}+t^{(fem)})/2-8\lambda_g,
\end{gather}
which is still {\em undetermined}. Therefore, $\lambda_g$ must be found to complete the derivation of the system of equations \eqref{5gre} and \eqref{10gre}.

In the general case of metrics $g_{ik}(x,u)$ this problem cannot be solved: a {\em complete} electricity equation (for $f_{ik,lm}$, not for the convolution \eqref{10gre}) is recently unknown. But it is
quite possible for the most important case of Weyl's conformal-Rieman-\ nian metric \cite{Weyl1918}.

\subsection{\textmd{Weyl's conformal metric}}
\label{Weylcl}
\quad Let us suppose as in \cite{Nskv2023} sec. 3.4.1, that Hermann Weyl was {\em right}: Maxwell's electrodynamics is a consequence of the scale invariance of the GR metric, that is equations \eqref{5gre},
\eqref{10gre}, \eqref{3geo} are satisfied by the conformal metric
\begin{gather}
  \label{1clw}
  g_{ik}=\chi(u)\,g^{(r)}_{ik}(x),\ \chi=\exp\left(2c^{-2}\beta\,A_lu^l\right),
\end{gather}
where $\beta=\rho/\mu=q/m=const$ \eqref{00as} (element of matter is a point particle) and $A_i$ is the $4d$ Maxwell vector potential.

In this case of {\em Embedding} linearity \eqref{4geo} of the operator $\hat b_{ik}$ gives
\begin{gather}
 \nonumber
  \qquad f_{ik,lm}=\hat b_{ik}g_{lm}=c^{-2}\beta\, F_{ik}\,g_{lm},  \\
  \nonumber
  \qquad F_{ik}=\partial A_k/\partial x^i-\partial A_i/\partial x^k,\\
  \nonumber
  \qquad f_{ik,lm}u^ku^lu^m=c^{-2}\beta\, F_{ik}u^k, \\
  \label{0clw}
  \qquad \quad 2f_{i \ \ \ \ \  l}^{\ [k,l]}=3c^{-2}\beta\, F_i^{\ k},
\end{gather}
and from $\xi_{ik}=9\varkappa_e\,F_{il}F_k^{\ l}/4\pi$ it follows that
\begin{gather}
  \nonumber
  \tau_{ik}=-t^{(fem)}_{\ \ \ \ ik}=18\,T^{(M)}_{\ \ ik},\quad t^{(oem)}_{\ \ \ \ ik}=-18\,t^{(L)}_{\ \ ik},\\
  \nonumber
  \qquad T^{(M)}_{\ \ ik}=\frac{1}{4\pi}\left(-F_{il}F_k^{\ l}+\frac{g_{ik}}{4}F_{lm}F^{lm}\right), \\
  \nonumber
  t^{(L)}_{\ \ ik}\equiv \frac{1}{4\pi}\left(F_iF_k-\frac{g_{ik}}{2}F_lF^l\right),\quad  F_i\equiv F_{ik}u^k,\\
  \label{2clw}
  \qquad \tau=-t^{(fem)}=0,\quad t^{(L)}=-F_iF^i/4\pi,
\end{gather}
where $T^{(M)}_{\ \ ik}$ and $ t^{(L)}_{\ \ ik}$ are the Maxwell and incom-\ plete  (see below) Lorentzian EMTs of the electro-\ magnetic field.

It is seen that the third line of \eqref{0clw} gives the {\em correct} expression for the Lorentzian acceleration, and the fourth line dictates the {\em choice} of the Lagran-\ gian multiplier $\lambda_g$, such
that the right-hand side of \eqref{10gre} is equal to $-3\Lambda_0 u_i$, which ensures the formal {\em coincidence} of \eqref{10gre} with Max-\ well's equation
\begin{gather}
  \label{02clw}
  \left[\left(-\Lambda_0+R/2\varkappa_g\right)g_i^k+\tau_i^k\right]u_k=-3\Lambda_0 u_i\Rightarrow\\
  \nonumber
  \qquad\qquad R/2\varkappa_g=-\left(2\Lambda_0+\tau uu\right).
\end{gather}
Then in view of \eqref{11gre} and \eqref{2clw} we find that
\begin{gather}
 \label{3clw}
  4\lambda_g=(5/4)\Lambda_0+(9/2)\left(t^{(L)}+T^{(M)}_{\ \ ik}u^iu^k\right).
\end{gather}

Using the result of \eqref{3clw} in the field equations, we can easily obtain their mutually consistent forms for the metric \eqref{1clw}:

-- the Maxwell-type equation\eqref{10gre}
\begin{gather}
 \label{4clw}
 \qquad\qquad \frac{\partial F_i^{\ k}}{\partial x^k}=\frac{\varkappa_e c^2}{2\beta}\Lambda_0\,u_i,
\end{gather}
which can be written in the standard form\footnote{$F_i^{\ k}\,\partial \ln\sqrt{-g}/\partial x^k\equiv0$ since $\hat b_{ik}$ and $\partial/\partial x^i$ belong to the {\em different} sets $M_c$ and $M_s$, \eqref{4geoo}.}
\begin{gather}
  \label{5clw}
  \qquad\qquad \frac{\partial(\sqrt{-g}\,F_i^{\ k})}{\sqrt{-g}\,\partial x^k}=-\frac{4\pi}{c}\, j_i
\end{gather}
where by virtue of \eqref{00as}, \eqref{0as}, \eqref{2as} we have
\begin{gather}
   \label{5clw*}
  \qquad\qquad j_i=\rho c dx^i/\sqrt{g_{00}}\,dx^0,
\end{gather}
which is an electric current density;

-- the Einstein-type equation for charged mat-\ ter with electromagnetic fields
\begin{gather}
   \label{6clw}
  \qquad R_{ik}-g_{ik}R/2=\varkappa_g\left(t^{(0)}_{\ \ ik}+t^{(d)}_{\ \ ik}\right),
\end{gather}
where
\begin{gather}
   \label{6clw@}
   \qquad\qquad\qquad t^{(d)}_{\ \ ik}\equiv t^{(dm)}_{\ \ ik}+t^{(df)}_{\ \ ik},\\
    \label{6clw*}
   \qquad \qquad\qquad t^{(dm)}_{\ \ ik}\equiv (5\Lambda_0/4)g_{ik},\\
  \label{6clw**}
    \!\!t^{(df)}_{\ \ ik}\equiv \! -18\!\left(T^{(L)}_{\ \ ik}+T^{(M)}_{\ \ ik}-\frac{g_{ik}}{4}T^{(M)}_{\ \ lm}u^lu^m\right).
\end{gather}
Here, $t^{(dm)}_{\ \ ik}$ is the EMT of dark matter, $t^{(df)}_{\ \ ik}$ is the EMT of dark energy and $T^{(L)}_{\ \ ik}$ is the {\em full} Lorentz EMT (with account of \eqref{3clw}) :
\begin{gather}
  \label{7clw}
    T^{(L)}_{\ \ ik}\equiv \frac{1}{4\pi}\!\left(\!F_iF_k-\!\frac{g_{ik}}{4}F_lF^l\!\right),\quad T^{(L)}=0.
 \end{gather}

\subsection{\textmd{Gravitational sources of Weyl's\\ {\em Embedding} }}
\label{dark}
\quad The resulting system of equations \eqref{5clw} and \eqref{6clw} of conformal {\em Embedding} is interesting first of all by gravitational-field sources due to the Lagrangian multiplier $\lambda_g$ of the second
equation. This multiplier has the meaning of the cosmologi-\ cal term of Einste-\ in's equation, $\Lambda_c=-4\varkappa_g\lambda_g$ \cite{Vlad2009}, and is usually associated with dark matter and dark energy
of the Universe. Therefore the new sources of the gravitational equation are combined in the EMT $t^{(d)}_{\ \ ik}$ and designated accord-\ ing to their matter-field content as $t^{(dm)}_{\ \ ik}\sim \Lambda_0$
and $t^{(df)}_{\ \ ik}\sim T^{(M)}_{\ \ ik}\&T^{(L)}_{\ \ ik}$.

The relation of dark sources $t^{(d)}_{\ \ ik}$ to ordinary gravity can be found by deriving $R_0^{\ 0}$-component of the Ricci tensor in the {\em Newton} approximation from \eqref{6clw}. In this case.
$u^i\simeq(1,\vec 0)$, and the only of $g^{(r)}_{ik}(x)$ metric component $g^{(r)}_{00}(x)\simeq 1+2\varphi_g$ that is different from the Minkowski value, gives, as we know,
$R_0^{\ 0}\simeq c^{-2}\triangle \varphi_g$, where $\varphi_g$ is the Newton potential, \cite{LL1988}. Then
\begin{gather}
  \nonumber
  c^2\triangle \varphi_g\simeq (\varkappa_g/2)\left[\mu_0c^2+\left(5\mu_0c^2/2-9\vec E^2/8\pi\right)\right]\Rightarrow\\
  \label{0drk}
  \ \triangle \varphi_g\simeq 4\pi k\left[\mu_0+\left(5\mu_0/2-9\vec E^2/8\pi c^2\right)\right],
 \end{gather}
where the terms in parentheses are dark gravitatio-\ nal sources.

It is seen that the contribution of "dark matter"\ (matter pressure) has the {\em gravitational} meaning, and the contribution of "dark energy"\ (electromag-\ netic fields) has {\em anti-gravitational} meaning.
Quali-\ tatively, the above result {\em correlates} with the interp-\ retation of astronomical observations {\cite{Riess1998,Perlmut1999,Zwicky1937}.

A quantitative estimate of the gravitational source composition can be found by comparing the summand moduli of the \eqref{6clw} right-hand side. For this purpose, let us represent the right-hand side
in an explicit form
\begin{gather}
  \nonumber
  R_{ik}-\frac{g_{ik}}{2}R=\varkappa_g\left[\left(u_iu_k-\frac54\ g_{ik}\right)(-\Lambda_0)+\right.\\
  \label{1drk}
  \quad \left.+\frac{9}{2\pi}\left(F_iF_k-F_{il}F_k^{\ l}+3\,\frac{g_{ik}}{16}F_{lm}F^{lm}\right)\right],
\end{gather}
and find the moduli of all summands averaged over the 3-volume on the assumption of “cold”\ $\vec v/c \simeq 0$ and at the “average almost flat”\ modern Universe
\begin{gather*}
  |\bar\mu_0 c^2|;\ 5|\bar\mu_0 c^2|;\ \frac{9}{2\pi}\left(|\,\overline{F^2}\,|-\frac{|\,\overline{F_{lm}F^{lm}}\,|}{4}\right)\!\simeq \!18 \bar w_e,
\end{gather*}
where $\bar w_e=\overline{\vec{E}^2}/8\pi$ is the average energy density of the electro-magnetic field.

Assuming that the principle of detailed equilibri-\ um $\bar\mu_0 c^2\simeq \bar w_e$ was {\em fulfilled} in the matter-radiation separation epoch and is still being fulfilled at present, it is easy to find the relative
weights of {\em visible} matter $n^{(vm)}$, dark matter $n^{(dm)}$ and dark energy $n^{(de)}$
\begin{gather}
 \nonumber
 \qquad n^{(vm)}=1/24\simeq4,2\%,\\
 \nonumber
 \qquad n^{(dm)}=5/24\simeq20,8\%,\\
 \label{2drk}
 \qquad n^{(de)}=18/24=75\%,
\end{gather}
-- which agrees with the well-known estimates of Universe's material composition\footnote{The comparison with \eqref{0drk} should not be confusing: gravitation is described not only by the
$\mathstrut_0^{\ 0}$-equation from the tensor block \eqref{6clw}, which for this evaluation must be used as a {\em whole}....}

The present-day spatial distribution of visible, dark matter and energy is the result of Universe evolution, in which stellar radiation plays a promin-\ ent part. It's quite possible that it “sweeps”\ some of the
visible matter to the periphery of star clusters, so that near the stars (on the star system scales) the dark sources compensate each other....

The obtained results \eqref{0drk}, \eqref{2drk} have a special meaning for the MES and the developed geometri-\ zation: they are a direct consequence of its structu-\ re and dimensionality. Hence, it is quite obvious
(coefficient $n-1=3$ \eqref{02clw}) that for a different dimension of {\em Embedding} the composition of Uni-\ verse would be different. Furthermore, since the pressure of {\em visible} matter is now identified
as {\em dark} matter, the problem of its detecting is merely {\em absent}.

\subsection{\textmd{Electricity and gravity: mutual\\ calibrations}}
\label{EGcali}
\quad In the general case of {\em Embedding's} metrics, the electric sector is specified by the {\em non-tensor} quantity $f_{ik,lm} $ \eqref{9geo} defined on $M_c$ \eqref{4geoo}
\begin{gather}
 \nonumber
  \quad f_{ik,lm}=\hat b_{ik}g_{lm}\equiv\partial c_{k,lm}/\partial x^i-\partial c_{i,lm}/\partial x^k,\\
  \label{1Ecali}
  \qquad c_{i,kl}\equiv(1/2)\,\partial g_{kl}/\partial u^i,\quad c_{i,kl}=c_{i,lk}
\end{gather}
for which $c_{i,kl}$ plays the role of a potential. It is easily seen that $c_{i,kl}$ is defined to within a gauge transformation \cite{Konopl1972}
\begin{gather}
  \label{2Ecali}
   \qquad c_{i,kl} \rightarrow c_{i,kl}+\partial \nu_{kl}/\partial x^i,\quad \nu_{ik}=\nu_{ki},
\end{gather}
where {\em tensor} $\nu_{kl}(x)$ is an arbitrary smooth coordi-\ nate function such that $\partial \nu_{ik}/\partial x^l\in M_s$ by virtue of \eqref{4geoo}. In the special case, $\nu_{ik}$ can be the
{\em Rieman-\ nian metric} $g_{ik}(x)$ or, as in the anisotropic gravity Einstein-type equation \eqref{7clw}, is the metric tensor $g_{ik}(x,u=const)$ of the  connectivity $\gamma_{i,kl}$ \eqref{7geo}
of the set $M_s$.

Hence, first of all, the anisotropic field $g_{ik}(x,\ \\u=const)$ can be viewed as the {\em gauge} field of the potential $c_{i,kl}$. That is, each of the $10$ different components of the tensor $g_{ik}(x,u=const)$
can be formally considered as a gauge field with the group $U(1)_{ik}$ of the potential $c_{l,ik}$ -- by analogy to the case of Maxwell's vector potential. Apparently, the constants of these one-parameter groups
also constitute a tensor. Secondly, the property of the metric $g_{ik}(x,u=const)$ of being a gauge field of the potential $c_{i,kl}(x,u)$ demonstrates the symmetric {\em independence} of the electric and
gravi-\ tational sectors of the theory, in particular, the possibility of {\em gravitationally independent} definition of the “cosmological”\ multiplier $\lambda_g$.

On the other hand, the case of {\em Embedding's} conformal metric \eqref{1clw}  allows one to easily show that the multiplier $\chi(u)$ is the gauge field of the gravitational potential $g_{ik}(x)$. Indeed,
the existence area of the function $\chi(u)$ is defined by the field $u$: $\chi(u)\in M_c$. Therefore, $\chi(u)$ and the operator $\partial/\partial x^i$ \eqref{4geoo} {\em does not} have a common domain
of existence
\begin{gather}
 \nonumber
 \qquad \chi(u)\in M_c,\quad \partial/\partial x^i \in M_s\quad \Rightarrow \\
 \label{1Gcali}
 \qquad\qquad\partial \chi(u)/\partial x^i =0,
\end{gather}
that is, the function $\chi(u)$ behaves as $const(x)$ at any point of $M_s$. And this fact provides invariance of the Einstein gravity of $M_s$ with respect to the transformation
\begin{gather}
 \label{2Gcali}
  \qquad g_{ik}(x)\quad\Rightarrow\quad \chi(u)g_{ik}(x),
\end{gather}
called the Weyl's calibration. This transformation {\bf generates} a new space-time structure -- {\em Embed-\ ding} $M_e$ with metric \eqref{1clw}, which has Maxwell-type electrodynamics \eqref{5clw},
and anisotropic Einste-\ in-type gra-\ vity \eqref{6clw}: $\chi(u)$ allows one to calibrate $g_{ik}(x)$ along the $u$-directions at any coordinate point $(x)$\footnote{Moreover, for the {\em general} case of
{\em Embedding's} metrics, there exists a generalization of the calibration \eqref{2Gcali} for each $_{ik}$-component of the metric: $g_{ik}(x)\Rightarrow\chi_{(ik)}(u)g_{ik}(x)$ with 16 vector potentials
$A^{(ik)}\,_l$, of which only 10 are independent, see e.g., Eqs. \cite{Konopl1972}, p. 51.}.

A similar mechanism accompanies the calibrati-\ on of \eqref{2Ecali}: since $\nu_{kl}=\nu_{kl}(x)$, $\nu_{kl}\notin M_c$ and therefore \eqref{2Ecali} gives {\em Embedding} $M_e$ with metric
\eqref{1clw} and the anisotropic gravity due to {\em anisotropic} embedding of the $M_s$ subset.

Thus, it can be considered that in the case of Weyl's {\em Embedding} the gravitational $g_{ik}$ and electric $A_i$ (“reduced”$c_{i,kl}$) potentials are the {\em mu-\ tual} gauge fields, and the mechanism
for realizing the gauge properties of {\em Embedding} differs markedly from the standard (single manifold) one for the obvious reas-\ on: the manifold $M_e$ ({\em Embedding}) even in the field approximation
is the embedding of {\em two} independent sets $M_s$ and $M_c$.

\section{\textmd{CONCLUSION}}
\label{concl}
\quad A consequent variant of the classical field geo-\ metrization, which is based on the new 4d-model of space-time (MES, {\em Embedding}) should be consi-\ dered as the principal result of the work. The model,
endowing the space with a local {\em anisotropic} structure, provides the appearance of {\em new} physi-\ cally significant information, which in principle cannot be obtained within isotropic continuum models.

In particular, the anisotropy of {\em Embedding} being the cause of the electromagnetism, is also manifested in its gravity. At the same time, the anisotropy of the Universe matter distribution {\em astronomically
observed} on large cosmological sca-\ les can be interpreted as a result of evolution of the anisotropic MES, and dark matter and energy can be interpreted as new (dark) gravity sources lying behind the
$\Lambda_c$-term of Einstein-type equations.

So far, the {\em complete} geometrization has been carried out only for the special case of {\em Embedding} with Weyl's conform-Riemannian metric. It is convenient because allows a linear restriction of the
MES electricity to the {\em vectorial} variant of Maxwell-type electrodynamics. This property all-\ ows us to find the undetermined Lagrange multip-\ lier $\lambda_g\sim \Lambda_c$ of the gravitational variational
field procedure, which removes the standard uncertai-\ nty from the sources of gravitational equation. The results are in good agreement with the recent quantitative estimates of the Universe composition: $1:5:18$
are the ratios of visible matter to dark matter and dark energy.

Furthermore, a) {\em “unobservability”} of dark mat-\ ter is obvious: the role of the latter is performed by the ordinary {\em pressure} of matter; b) {\em pressure} increases the {\em gravitational} action of
matter; c) dark energy (of electromagnetic-field nature) pro-\ duces  the {\em anti-gravitational (!)} action and, finally, d) all of the above mentioned factors are the direct consequence of 4-dimensionality of
the MES.

Comparison with the known gravitational fields of the Universe suggests mutual compensation of “dark”\ sources near the stars and their increasing unbalance at a distance. More accurate estimates
of the composition and spatial-temporal distribu-\ tion of matter in the Universe can be obtained by solving the evolution problem of the {\em real}, fully geometrized {\em Embedding}.

Indirect or astronomical arguments of the MES correctness are certainly important, but they cannot be compared with the results of direct {\em experimental} tests, such as “red shift”\ measure-\ ments in the
electromagnetic fields \cite{Nskv2023, Nskv2017}. This physical effect exists only in the MES, and its study can provide direct information about both the anisotropy and the relic charge of the Universe. For the
electrostatic field in vacuum the calcula-\ tion formula is $\Delta\omega/\omega\simeq \mp c^{-2}\sqrt{k}\,\Delta\varphi = 10^{-21}\Delta\varphi$, where $\Delta\varphi(V)$ is the difference (in volts) of the
photon track potentials, and the sign must be determined experimentally (depends on the sign of Universe relic charge). In our opinion there is an acute need for appropriate experiment: it is much cheaper than
modern gravitational measurements!

\medskip
The work is performed under support of RAS (№ 124012300246-9).
\medskip

\renewcommand{\refname}{

\end{document}